\documentclass[conference,10pt]{IEEEtran}
\IEEEoverridecommandlockouts
\usepackage{cite}
\usepackage{amsmath,amssymb,amsfonts}
\usepackage{algorithmic}
\usepackage{graphicx}
\usepackage{textcomp}
\usepackage{xcolor}
\usepackage{adjustbox}
\usepackage{array}
\usepackage{subcaption}

\newcolumntype{R}[2]{%
    >{\adjustbox{angle=#1,lap=\width-(#2)}\bgroup}%
    l%
    <{\egroup}%
}
\newcommand*\rot{\multicolumn{1}{R{20}{1em}}}% no optional argument here, please!

\def\BibTeX{{\rm B\kern-.05em{\sc i\kern-.025em b}\kern-.08em
    T\kern-.1667em\lower.7ex\hbox{E}\kern-.125emX}}
\begin{document}

\newcommand{\name}{\textsc{FLIC}}
\newcommand{\notenk}[1]{\textcolor{red}{\textbf{NK: #1}}}
\newcommand{\notegkt}[1]{\textcolor{blue}{\textbf{GKT: #1}}}

\title{\name: A Distributed Fog Cache for City-Scale Applications\\
}

\author{\IEEEauthorblockN{Jack West}
\IEEEauthorblockA{\textit{Department of Computer Science} \\
\textit{Loyola University Chicago}\\
Chicago, Illinois, USA \\
jwest1@luc.edu}
\and
\IEEEauthorblockN{Neil Klingensmith}
\IEEEauthorblockA{\textit{Department of Computer Science} \\
\textit{Loyola University Chicago}\\
Chicago, Illinois, USA \\
neil@cs.luc.edu}
\and
\IEEEauthorblockN{George K. Thiruvathukal}
\IEEEauthorblockA{\textit{Department of Computer Science} \\
\textit{Loyola University Chicago}\\
Chicago, Illinois, USA \\
gkt@cs.luc.edu}
}

\maketitle

\begin{abstract}

We present \name, a distributed software data caching framework for fogs that reduces network traffic and latency.
\name~ is targeted toward city-scale deployments of cooperative IoT devices in which each node gathers and shares data with surrounding devices.
As machine learning and other data processing techniques that require large volumes of training data are ported to low-cost and low-power IoT systems, we expect that data analysis will be moved away from the cloud. 
Separation from the cloud will reduce reliance on power-hungry centralized cloud-based infrastructure.
However, city-scale deployments of cooperative IoT devices often connect to the Internet with cellular service, in which service charges are proportional to network usage.
IoT system architects must be clever in order to keep costs down in these scenarios.
To reduce the network bandwidth required to operate city-scale deployments of cooperative IoT systems, \name~ implements a distributed cache on the IoT nodes in the fog.
\name~allows the IoT network to share its data without repetitively interacting with a simple cloud storage service, reducing calls out to a backing store.
Our results displayed a \textbf{less than 2\% miss rate on reads.}
Thus, allowing for only 5\% of requests needing the backing store.
We were also able to achieve \textbf{more than 50\% reduction in bytes transmitted per second.}
\end{abstract}

\section{Introduction}

IoT devices that are deployed in locations where internet connections are not available often rely on cellular service to provide connections to the cloud.
These devices support a wide range of infrastructure, including smart cities, automated vehicles, oil and gas production, and many more.
Some industrial building automation systems even use cellular connectivity for ease of deployment and management.

When cellular service provides connectivity to an IoT ecosystem's backend, ISPs typically charge by the byte---the more traffic a device generates, the higher the bill.
This is more or less the billing structure used for consumer cellular plans.
IoT system architects must be clever in order to keep costs down in these scenarios.

Managing network traffic is particularly challenging in distributed systems where nodes both produce and process sensor data---a network structure that is becoming increasingly popular given the high costs of maintaining centralized datacenters filled with GPU-enabled servers~\cite{videoanalytics,waggle}.
In systems where nodes serve as both producer and processor, network usage is often heavy as nodes construct system models from training data generated by their neighbors.
The data exchange is often done by way of the cloud backend, which provides storage and coordination for all the data and mutually authenticates the nodes.
The trouble with backend mediated storage is that the overall deployment incurs network load in proportion to the number of nodes.
While compressive sensing and other coding techniques can reduce network traffic, the storage hierarchy used by many applications uses a cloud-based manager to store data and coordinate actions in these distributed systems still require a large amount of network bandwidth and represent a single point of failure.

\name~ is a distributed cache architecture for fogs intended to reduce the network bandwidth and latency required to share large volumes of data among coordinated nodes.
\name~ is a layer that lives between the application code and the backend API that provides a distributed software cache for backend transactions, which makes it easy to incorporate into new or existing software architectures.
It is implemented as a transparent module so that software architects do not have to think about what is the best way to share data among many cooperative nodes in a fog.
Programmers can issue read, write, and update requests to \name~ in the same way as they would to a relational database backing store.

\name~ is built with the broad class of distributed machine learning systems in mind.
The application scenario we envision is one in which end nodes cooperatively collect and classify data in the field, and a relatively small portion of the work is outsourced to the cloud-based backend.
Many examples of coordinated computer vision systems have been published recently~\cite{videoanalytics,waggle,gauen2017comparison,cisco_fog_paper}.
Data sharing in such systems is normally coordinated by some sort of edge or cloud-based service.
In city-scale applications, devices are often connected to the cloud by way of a cellular modem, which introduces latency, bandwidth, and cost limitations.

The architecture of \name~ is designed to support such distributed systems in a data-agnostic way.
\name~ uses locality of reference to decide which data should be stored in the local distributed cache and which should be written out to the backend database.
Our locality of reference strategy performs well under the workloads we tested and is easily portable to different applications because it does not use domain-specific information about the data or the application.
Our assumption about the locality of reference in distributed computer vision applications is a mostly empirical one based on our experience working around these kinds of systems.

\subsection{Challenges}

Cache coherence and transaction timing are the two most important challenges of implementing a data-agnostic software cache for fogs.

\paragraph{Coherence}
\name~ distributes new data to nodes on the local network using UDP broadcasts.
Broadcast packets are occasionally dropped because of collisions and interference on the communication channel.
Our goal is to maintain coherence among many distributed nodes under the assumption that communication is unreliable.
This is complicated by the fact that the same row of data may be repetitively overwritten or updated without updating all copies in the fog.
%Maintaining \textbf{soft coherence} among many distributed nodes under the assumption that communication is unreliable is hard, particularly if the same row of data is repetitively overwritten or updated.
%This can cause some nodes to unknowingly store stale versions of data that should have been updated or overwritten.
We implement a kind of \textbf{soft coherence} in \name~ in which the system can tolerate a portion of the nodes storing stale versions of the data, as long as at least one node has the most recent update.
Soft cache coherence is a loss-tolerant cache coherence strategy which we discuss in detail in the architecture section.
%The more nodes you have, the more likely that you will have at least one copy of a given piece of data.
%So this doesn't work on small swarms because of the possibility of packet loss.
If, for example, a node requests an entry from the fog cache and gets multiple different data values back, it accepts the one with the most recent timestamp.
In our simulated implementation of \name, nodes are implememnted as containers on a common host, so they all have a common synchronized clock.
However, clock synchronization is not strictly necessary for correctness.
We discuss later how \name~ uses soft coherence to reduce the redundancy of saved data in the distributed cache.

\paragraph{Timing}
In real fogs and our evaluation of \name , reading, and writing data to the backing store is a slow process.
For the backing store we used in our evaluation, the write latency is longer than the arrival period of new data, a problem that gets worse as the number of nodes in the fog grows.
Individual nodes do not directly talk to the backing store because the API's latency is too long.
Instead, our implementation of \name~ uses a single queued writer task, similar to a CPU's load-store buffering~\cite{zerocycleloads}, to write data from all nodes in the fog to the backing store.

\subsection{Contributions}

Our main contributions are
\begin{itemize}
\item We build \name , a modular and data-agnostic fog-based software caching system that is designed to reduce the network bandwidth and latency for nodes within a fog.
\item We introduce the notion of soft cache coherence in which updates to a distributed cache may be lost. We evaluate soft coherence with a simulated workload.
\item We evaluate \name~ on a simulated fog network, demonstrating that it reduces network bandwidth by 50\% in our experimental workload.
\end{itemize}

\section{Architecture}

\name~ is targeted at distributed fog deploments in which multiple types end devices coexist on a common fog network.
We assume that each device interacts individually with a single cloud service.
The end devices may not all have the same functionality or process the same kind of data. 
We assume that they are all part of the same ecosystem that uses a single cloud-based database to archive information.
This is a standard configuration in building automation and personal fog deployments.

\begin{figure}[t]
    \centering
    \begin{subfigure}[b]{0.48\textwidth}
        \includegraphics[width=\textwidth]{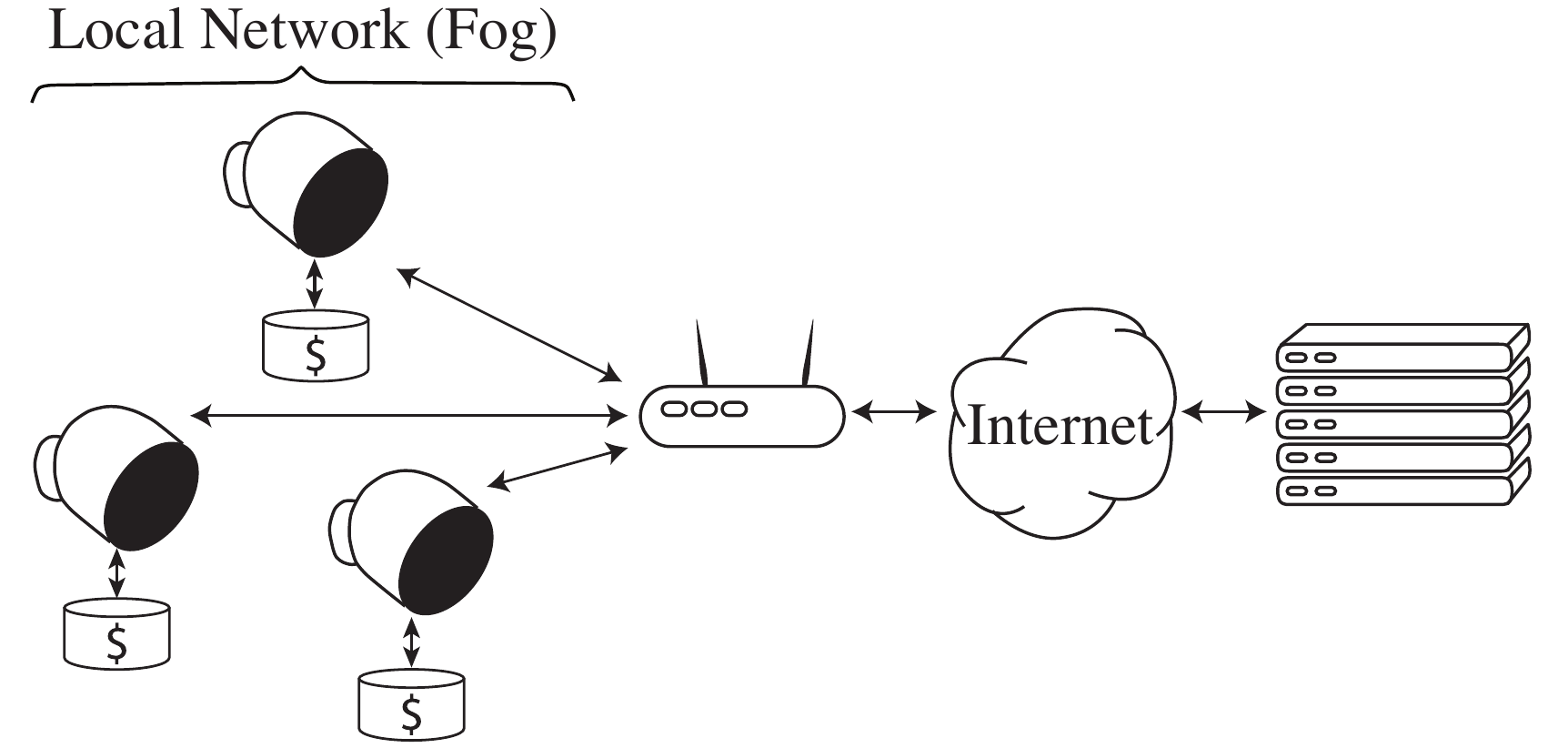}
        \caption{Organization of edge devices in a fog.}
        \label{fig:archdiagram}
    \end{subfigure}
    \begin{subfigure}[b]{0.48\textwidth}
        \includegraphics[width=\textwidth]{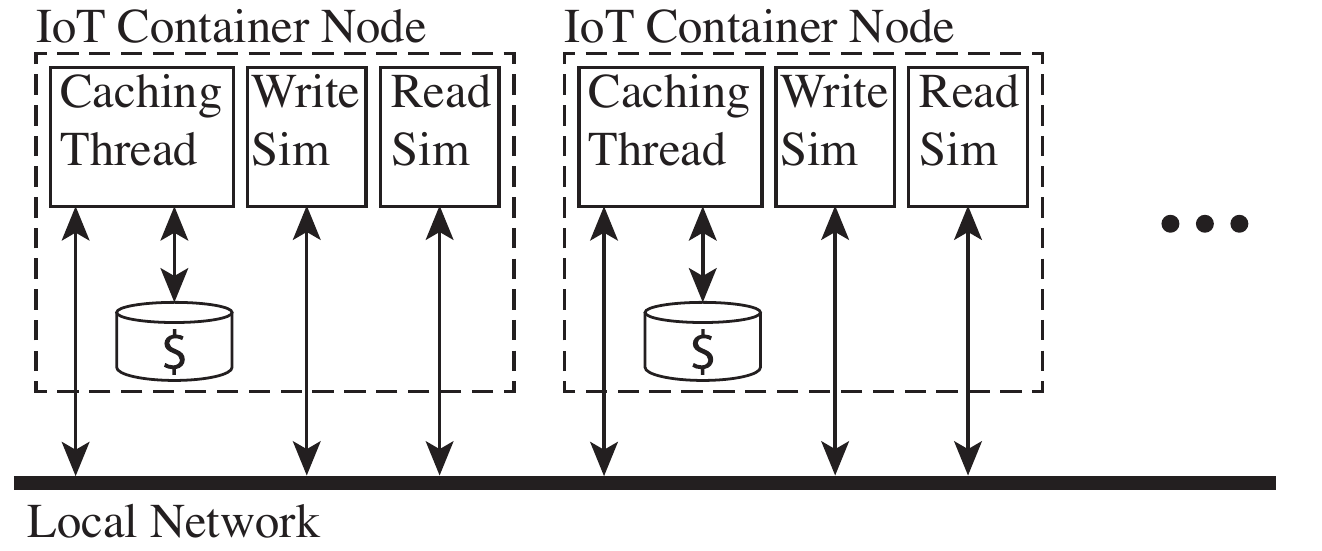}
        \caption{Software architecture of our fog cache.}
        \label{fig:softarch}
    \end{subfigure}
\caption{Distributed cache on a fog of devices.}
\vskip -15pt
\end{figure}

As individual devices in the local fog generate new data (for example from sensor readings, etc), it is stored locally in the distributed cache that is shared among the devices on the fog network.
If data needs to be read by one of the local devices, it can be taken first from the distributed cache.
If the required data is not in the distributed cache, the cloud-based database server, which serves as a backing store, will be queried.
Fig. \ref{fig:archdiagram} depicts the configuration of the system.

%\paragraph{Prototype Implementation}

We constructed a prototype implementation of a fog of devices using Docker containers.
We use Google Sheets as the backing store for our distributed cache because it has an accessible API, and it is a realistic system.
The system is built on two Python 3.7 scripts.
One of the scripts acts as the main operator for the data stream.
The other script acts as a router.
The first script is applied to each node that generates data and utilizes the global caching system.
The final node is built on top of the script that emulates the router; we wanted to have a single point of transmission when sending data over the wire.
Routing the data at a singular point allowed us to create a choke point to manage the data.
The router script properly managed the packets such that no two packets were sent at once.

\subsection{Node Architecture}

\begin{table}
\caption{Structure of the cache with example data.}
\center
\vskip -10pt
\begin{tabular}{llllll}
\rot{\textbf{Index}}  & \rot{\textbf{Valid?}} & \rot{\textbf{Time Inserted}} & \rot{\textbf{Data Timestamp}} & \rot{\textbf{Node ID}} & \rot{\textbf{Data}} \\ \hline
1 & 1 & 1568673296 & 1568673295.5 & 7 & XXX \\
2 & 1 & 1568673296.5 & 1568673295.5 & 5 & XYX \\
3 & 0 & 1568673295.25 & 1568673200 & 2 & ZZZ \\
4 & 1 & 1568673290.125 & 1568673289 & 3 & XYZ \\
5 & 1 & 1568673293.125 & 1568673290 & 7 & ZXY \\
\end{tabular}
\label{tab:cachestructure}
\vskip -10pt
\end{table}

%The miniature operating system's job is to generate data.
There are three major software components on each node that are used to implement the fog cache:

\begin{enumerate}
\item \textbf{Caching Simulator:} The caching thread manages the global cache specific to each node. 
It also distributes tasks based on packets received from the fog network.
This thread is also responsible for letting the backing store know of a key change that was generated from that node specifically.  
\item \textbf{Write Simulator:} The write simulator generates data and then multicasts the data on the Docker network to all nodes. Writes are done once per second.
\item \textbf{Read Simulator:} The read simulator asks the fog network for a specific key and value.
All requests for data are values that the cache has a record of existing.
Whichever nodes had the value requested is also kept track of in the read simulator.
\end{enumerate}

%All of these components are maintained on the node's operating system written in python.
Each of these components is implemented as a thread in python.
%Localizing the data is one of the issues that we set out to solve.
%With each node logging its data at run-time we can examine in real-time the functionality of the software stack.
The threads running on each node log timestampped events to a common text file which we use to generate plots in the evaluation section.
We simulated a fog of devices in Docker because it is a lightweight virtualization framework that allows us to simulate many nodes on a single machine and because it allows us to save event logs in one location.
%We decided to use Docker due to its customizability.
%Due to the fact that nodes of an IoT network can be so different yet have to work in harmony requires the need for customizablility.

Our prototype uses UDP multicast packets to share data on the local network.
All Docker nodes live on a shared local area network.
The fog connects to the Google Drive backing store by way of an HTTPS connection over TCP.
%Whereas, a TCP connection is forged with the Google backing store.
All traffic to the backing store is routed through a single Docker container.
This was done to deal with conflicted writes to Google Drive's API and also to reduce the number of calls to Google's API, which is limited to 500 calls per 100 seconds.
%Along with data management, the choke point of the topology is designed to limit one TCP connection per fog. 

\iffalse
The system is scalable.
We designed the system to be able to run with a mutable size for the fog.
We also decided to use Docker images similarly to Foggy\cite{foggy}.
Like Foggy, we took advantage of the simplicity of Docker as well as the customizability Docker provides.
Creating a fog network on top of Docker containers allows this experiment to also be re-creatable on any Linux system with Docker functionality.
\fi

% The Fog is also scalable.
% We designed the system to be able to add and substract nodes whenever the user deciedes to.
% The user would just have to generate a new image and map that new image to the node.
% Multiple images can also be applied to the same node.
% Thus, if the user wants to have several nodes that operate on the same software stack that is also possible.
% The dynamic nature of the system needed to also work for a network that wanted to experiment as well.
% Docker virtual containers are put on their own network so monitoring their traffic is simple.

\subsection{Soft Cache Coherence}
In a fog that connected by some sort of wireless network, we assume that the communication channel used to update entries in a distributed cache may be subject to losses.
\name~ uses UDP broadcasts on the local area network to update cache lines on neighboring nodes.
These messages can be lost to network interference, queue overflows, etc.
Soft cache coherence is the term we use to describe loss tolerance.
When one node broadcasts a row update to the surrounding fog, that update will be recorded by all nodes that can hear the broadcast.
Since each row (in our evaluation) is timestamped, we will be able to determine what the most recently updated data is in case of a disagreement among nodes.
Soft coherence cannot guarantee the integrity of data in the cache, but we can bound the probability of error parameterized on the loss rate of the network.
%We omit that analysis here because of space limitations.

Soft coherence requires broadcast packets to work.
In our implementation, we used broadcasts on a local area network, but this is not necessarily required.
It would be possible to apply soft cache coherence in fogs where nodes are not all on a common subnet as long as there is some mechanism in place to support broadcasts, for example a VPN or VLAN.

To get a general idea about the loss probability under soft cache coherence, we can calculate the probability that a broadcast is completely lost---this is the probability that it is lost at every node in the local network:

\begin{equation*}
Pr[Loss] = Pr[L_1 \cap L_2 \cap L_3 \cap ... \cap L_N]
\end{equation*}

\noindent
where $L_k$ is the event that the broadcast packet is lost at node $k$.
We can think of the $L_k$'s as Bernoulli random variables that take the value of 1 in the event that a broadcast packet is lost at node $k$ and the value of 0 if the packet is correctly received.
To get the number of nodes at which a broadcast packet is lost, we can calculate the sum of the $L_k$'s for all nodes in the fog.
Applying the Markov inequality:

\begin{equation*}
Pr[Loss] = Pr[\sum L_k \geq N-1] \leq \frac{E[L_k]}{N-1}
\end{equation*}

Where $E[L_k]$ is the mean of the $L_k$ random variable, the same as the probability of a loss at one node.
As the number of nodes in a fog increases, the probability of a complete loss decreases under soft cache coherence.

Practically speaking, we can say that some relatively small fraction of the data collected by the fog will be lost.
We assume that in the applications that are likely to use \name , individual cache lines are not mission-critical by themselves.
Losing one cache line will not be catastrophic for the application as long as most of the collected data is recorded.
This is because we are assuming that the data stored in the cache will be used to train a machine learning model, which captures larger trends.
In other words, we assume that there is some amount of redundancy in the cached data.

\subsection{Data Compression}
In our prototype implementation, each node in the fog generated uniformly distributed random data to be cached, which had similar statistical properties to data that had been encrypted and compressed.
\name~ adds another layer on top of compression, which can further reduce the network bandwidth needed to store and share the data among nodes in a fog.

%We wanted to do this such that fog topologies which are location dependent can get better results by comparing what they recieved from other nodes.
%This shared computation allows more confident results to be recieved directly from the fog rather than doing all computation on a single cloud or a single server.
%More accurate data directly from a fog network would allow for the user to take less time and energy parsing and analyzing results from a network.

\subsection{Backing Store}

In our prototype implementation, we use Google Sheets as our backing store for data that does not fit in the fog-based cache.
The cache stores a subset of the data generated in the fog, and old rows are evicted to the backing store as new rows are generated.
We evict the least recently used cache lines.
Google Sheets is a reasonable choice for a backend because it is a widely used platform implemented by a professional development team.
It is also the heir apparent to Google Fusion Tables, which provided a large-scale table framework for tabular data. This technology has been merged into Google Sheets.
We did not want to implement a custom backend service for \name 's prototype implementation because the properties of a custom implementation could affect the results of our evaluation.
We also wanted a proper cloud to live on top of our fog to better represent the definition of fog computing\cite{fogcomputingapps}.

Google sheets does have some drawbacks, some of which have analogs in realistic cloud-based database systems.

\textbf{Rows that arrive contemporaneously overwrite each other.}
In a custom backend implementation that uses a database as a backing store, we would normally save the contents of both rows or retain the data from the most recent row only.
Because Google Sheets is not implemented as a relational database, it is not transactional.

\textbf{Latency to backend services tends to be high.}
This is an unavoidable problem, even in bespoke cloud-based database systems, because servers need to support large deployments of edge devices.
Much of the latency in this system comes from the RESTful API, which uses HTTPS.
Realistic cloud-based backends will all suffer from high latency to a greater or lesser extent.
One of the main goals of \name~ is to hide this latency from the application running on the end device.

Fog topologies try to reduce backend latency by doing more computation on end devices.
As we demonstrate in Section \ref{sec:eval}, reading from Google's backend will only increase as more data is saved on the backing store.
This is an artifact of the Google API---we cannot query a Google Sheet for specific data using the AppScripts API provided by Google.
Instead, in order to read, the end device must grab the entire contents of the Google Sheet and parse it locally, searching for the desired row.
This feature of Google Sheets is probably not common to most cloud-based backends, and it causes significantly higher network traffic when we have to read from the backing store---for example when \name~ misses on a read.

\textbf{Google imposes a limit of 500 API calls per 100 seconds~\cite{Google_Sheets_Quota}.}
This hard limit imposed by our backing store creates conditions similar to spurious connectivity issues in real network deployments.
It also imposes throughput limitations on our simulated app.
%If one were to reach the limit, the python API would raise an error.
%Fixing this problem inadvertently attributed to us creating our fault tolerance system.
When the backing store fails, the cache thread queues incoming write requests locally and reattempts the write using a backoff strategy similar to binary exponential backoff used by Ethernet channel access.
Meanwhile, the data is available to be read by other nodes in the fog.

%Google's API does have limitations with how many calls a user is allowed to make per project. The limit is 500 calls per every 100 seconds. The limitation halts some aspects of scalability. The more IOT devices a user has, the bigger the local temporary cache will have to be. 

%All tests have been done in a virtual environment. The aim of the tests have been to emulate an IOT network as close as we could. The issue is that no IOT network has implemented the system for use. 

\section{Evaluation}
\label{sec:eval}

\subsection{Evaluation Methodology}

We are evaluating \name~ on a simulated fog network where edge devices are all generating sensor data and sharing that data to make some statistical inferences.
Our simulation platform is a fog of Docker nodes on a Linux server with 16 GBytes of RAM and an Intel Core i7 CPU.
When we ran simulations, we only measured network traffic generated by the simulated fog network.

\subsection{Workload}
\name~ was designed with a fog-based data collection and classification system in mind, and, like CPU memory caches, we expect its performance to be dependent on the workload.
The workload we chose in this evaluation is consistent with the use case we envision for it.

The kind of workload we are trying to emulate is one in which the fog consists of many sensors that are collecting data in real-time and logging that data to their cloud service.
Waggle~\cite{waggle,cityscale} and other city-scale applications use this type of structure for data collection and analysis.
We assume that the nodes are doing some data processing locally and using cloud services as a large backing store for historical data.
We assume that older data is less critical.
%We are thinking about machine learning at the edge that needs large volumes of recent data to be available.
This is a typical architecture for smart city applications in which nodes need cellular network connections for internet access.

In our workload, each node writes new data to the cache every second.
Each node also randomly reads data every 15 seconds.
Random read keys are chosen from the node's global cache.
Once a key is chosen, the node sends out a request to other nodes in a fog.
Reads are done in this way because it emulates the behavior of preferentially reading recent data.

The cache sizes on each node represent the number of entries that the cache is able to hold.
For example, one node in a fog of distributed security cameras may only have enough memory to store 200 images in its cache.
%If one has a node that only has enough data for 200 pictures on the device's ram, i.e the user wants to think in data entries rather than byte sizes.
%To think in data entries allows the system to be applied to any type of data or size limitation.

\subsection{Responsiveness}

\begin{figure}[t]
\centering
\includegraphics[width=0.45\textwidth]{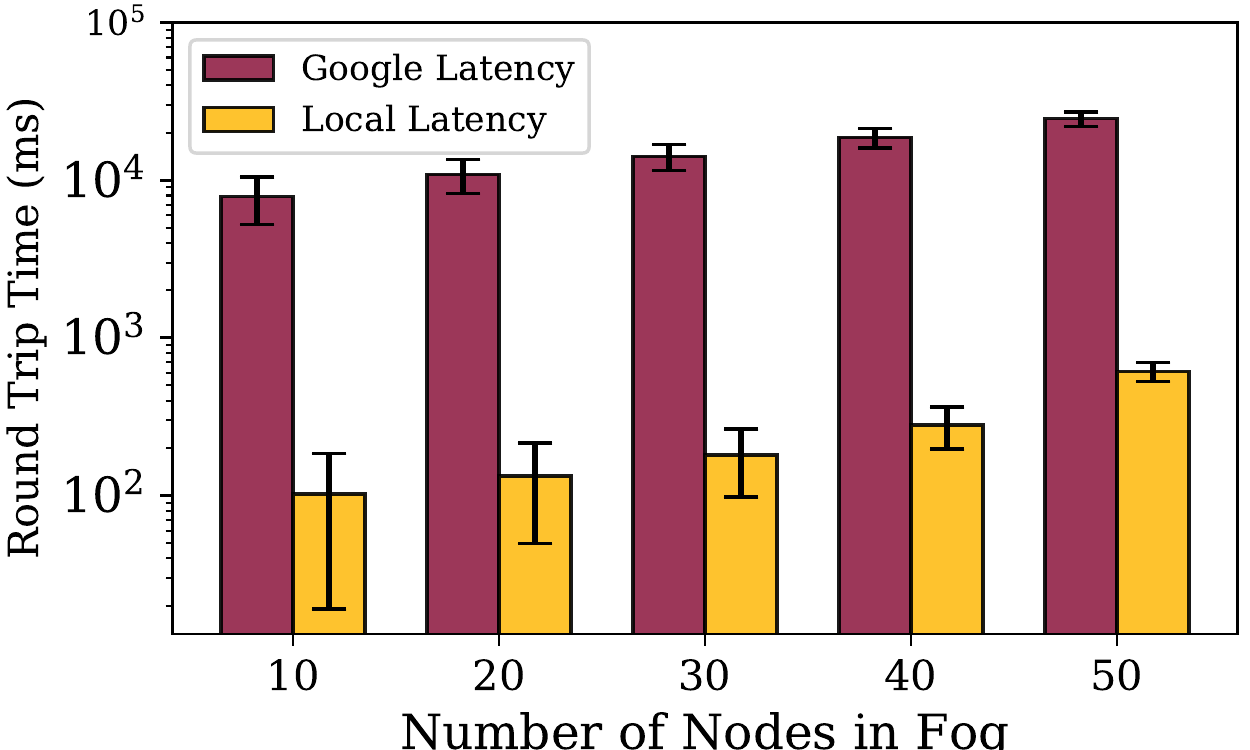}
%\vskip -10pt
\caption{Communication latency in a fog. Note: log scale.}
\vskip -15pt
\label{fig:rtt}
\end{figure}

In this experiment, we test the round trip time (a) from one node to the Google Sheets backing store and (b) from one node to all other nodes in the fog.
Fig. \ref{fig:rtt} shows latency in each case for different fog sizes.
On the x-axis, we vary the number of nodes in the fog, and on the y-axis, we measure the round trip time to the backing store and to all other nodes in the fog (with a logarithmic scale).
Like in our cache implementation, we use broadcast packets to measure latency, and we measure response time as the amount of time it takes for all nodes to respond to a broadcast.
Because we are evaluating \name~ on a simulation composed of a fog of Docker nodes, we think that the increase in latency to other nodes in the fog is probably caused at least in part by a bottleneck in CPU cycles on the evaluation machine.
In a real deployment, the individual nodes would not have to compete with one another for CPU time, and we expect that the latency would not increase as dramatically.

\subsection{Bandwidth}

\begin{figure}[t]
\centering
\includegraphics[width=0.45\textwidth]{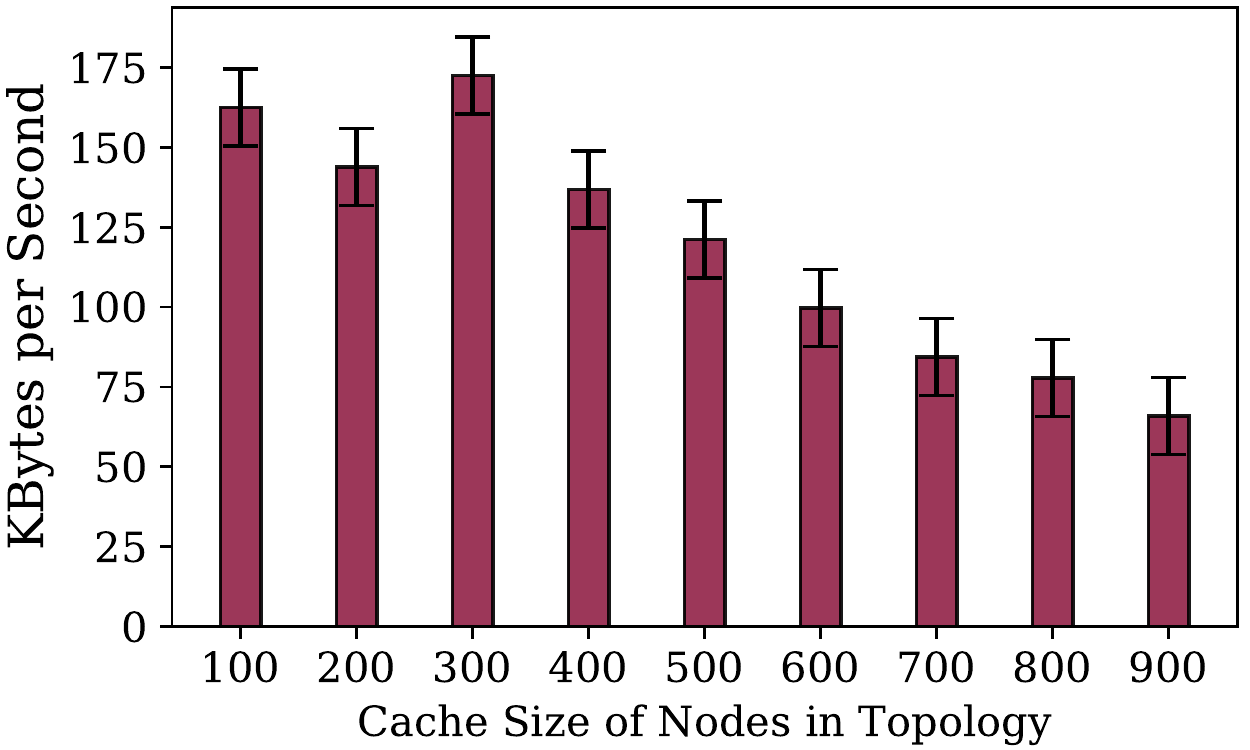}
%\vskip -10pt
\caption{Data rate on wide area network (x-axis)  as a function of number of nodes in the fog (y-axis).}
\label{fig:rowsize}
\end{figure}

In this experiment, we measure the amount of data transmitted or received per second on the WAN as a function of cache size.
Because in cellular networks, ISPs charge per transmitted byte on the WAN, we are interested in measuring \name 's impact on data on the WAN.
Fig. \ref{fig:rowsize} shows the number of bytes transmitted per second on the y-axis as a function of the cache size (x-axis).
We fixed the number of nodes in this experiment at 50 nodes.
As we would expect, network traffic goes down as cache sizes increase because reads hit more often in the cache.

In steady-state, as nodes generate new data---in our case once every second---\name~ will not be able to hide latency or reduce the bandwidth of writing out to the backing store because each new data element will evict an existing data element that is already in the cache.
The primary source of gains in this experiment is reads.
Reads that hit will not need to access the backing store at all, so increasing the read hit ratio is the best way to reduce network traffic.
The ratio of reads to writes---a parameter that depends on workload---will dictate performance gains in this experiment.
We chose a conservative ratio of 15 to 1 for our evaluation.
Real workloads will likely have a more even ratio of reads to writes, which would significantly improve the performance of \name.

\subsection{Read Miss Ratio}

\begin{figure}[t]
\centering
\includegraphics[width=0.45\textwidth]{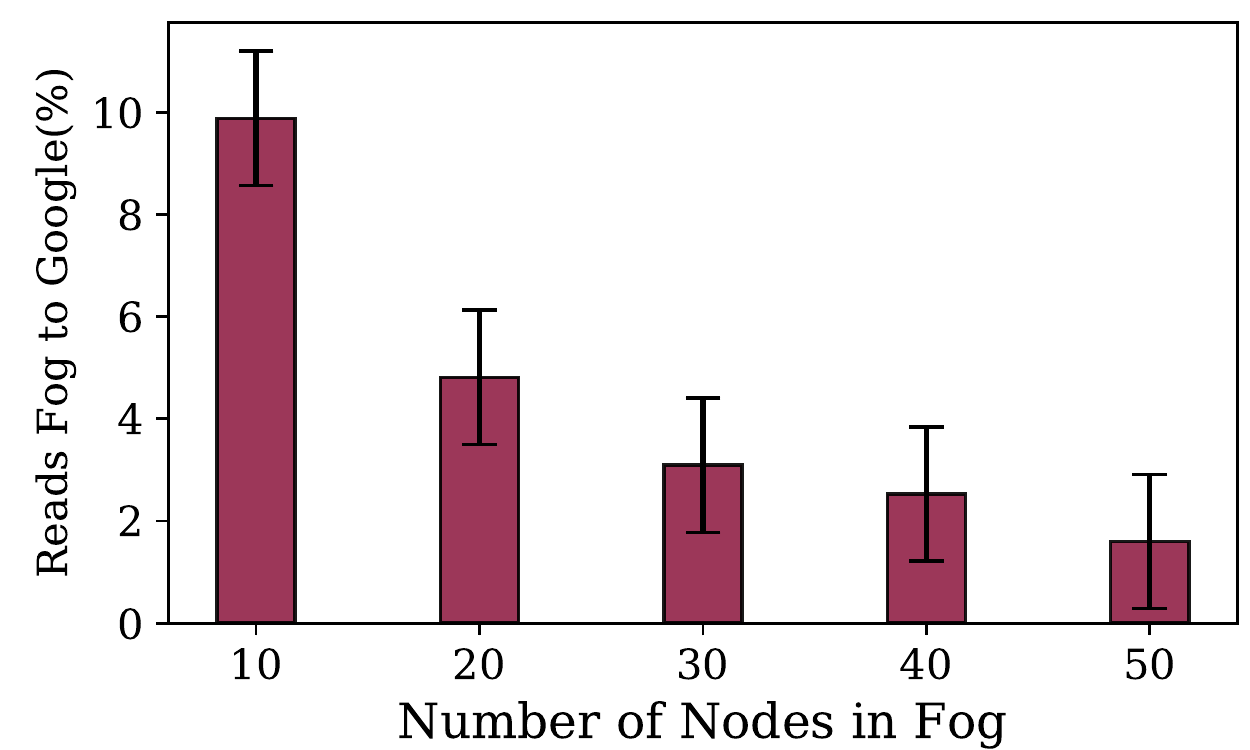}
%\vskip -10pt
\caption{Fraction of reads that miss on the distributed cache.}
\vskip -15pt
\label{fig:missratio}
\end{figure}

In this experiment, we test how the read miss ratio is affected by the number of nodes in the fog.
Our goal is to make the read miss ratio as close to zero as possible, which reduces the total network traffic to the backing store.
Fig. \ref{fig:missratio} shows the read miss ratio (y-axis) as a function of the fog size (x-axis) for a fixed cache size of 200 lines.
The miss ratio goes down significantly as the fog size increases, as we would expect because of the total amount of available storage in the fog-based cache increases.

\subsection{Transaction Sizes}

\begin{figure}[t]
\centering
\includegraphics[width=0.45\textwidth]{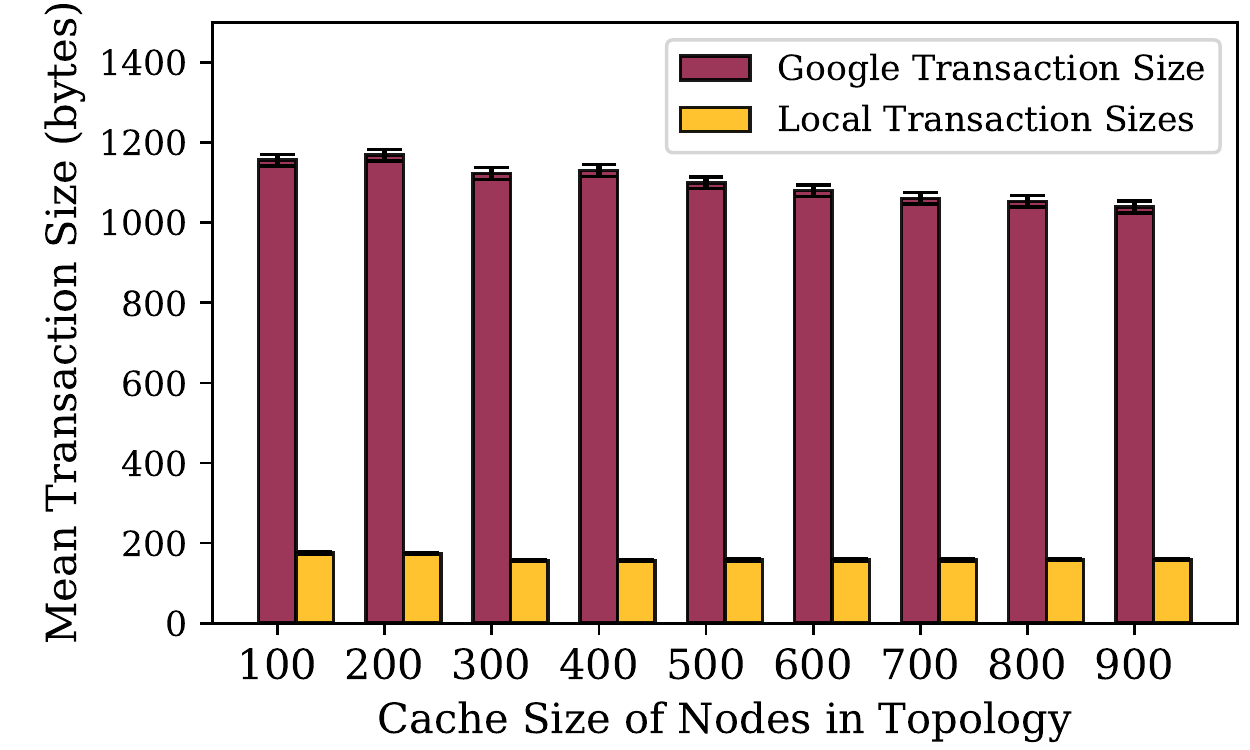}
%\vskip -10pt
\caption{Average size of transaction.}
\vskip -15pt
\label{fig:transactionsize}
\end{figure}

In this experiment, we measured the average size of transactions issued to the backing store.
Fig. \ref{fig:transactionsize} shows the average transaction size to the backing store (y-axis) as a function of cache size (x-axis) for a fog consisting of 50 nodes.
As the cache size increases, the read miss rate decreases.
This causes transactions to the backing store to be slightly smaller overall.
With more data to be used on each node's global cache, there are fewer calls required to go out of the system.
Google requires we pull the entire database upon each read.
As the database fills, more data is pulled upon each read.
So by \ref{fig:missratio}, we see that fewer calls are made to Google as the cache sizes grow.
So fewer calls to Google lowers the average transaction sizes.
To further this claim, there is also a slight trend upwards on the local transaction sizes.
It is hard to see on the graph, but on the same data points that show a downward trend on Google, there exists an upward trend on the local transaction sizes.

\section{Discussion}

We used the concept of soft cache coherence when implementing \name.
Soft coherence is the concept that the distributed cache can maintain a faithful representation of the data it stores even if the communication channel used to maintain cache coherence is unreliable.
Under the assumption of unreliability, the probability that the most recent update of a row is available somewhere in the cache is proportional to the (1) the probability of loss on the communication channel and (2) the number of nodes in the fog.
We will leave a full exploration of the topic to future work.
\iffalse
From the evaluation section, we can suffice that there is a trend for fewer bytes per second.
Fewer bytes per second allows an fog network to be more affordable and approachable for experimentation.
Thus, furthering experimentation and interest into fog computational networks.
We also see the decrease in transaction sizes, which better supports that more reads are made locally.
Thus, allowing for faster processing among the fog.
\fi
%The wise know, to not let the perfect be the enemy of the good.

\paragraph{Time Synchronization}

\name~ does not require all nodes in the fog to have synchronized system times.
In our implementation, all nodes did have synchronized system times because they were all running as containers on the same host.
The key that we use to store lines in the cache is generated from a hash of a long string that includes the timestamp at which the data was generated.
But because that generation timestamp is included in the cache line, it is not necessary for the individual nodes to all agree on the system time.
To compute the hash, they just need to look at the timestamp field within the cache line.

\paragraph{Shortcomings}
Although we demonstrated that \name~ can improve responsiveness and reduce network traffic on the wide area network, it does have shortcomings.
First, \name 's performance is closely tied to workload.
Its performance will be a function of the kind of data being exchanged in the fog and the ratio of reads to writes.
In workloads where nodes only write data to the backing store and do not read, \name~ will not improve performance because each new write will cause a cache eviction and trigger a write to the backing store.
However, workloads with relative parity between read and write operations will experience performance improvements.
In our evaluation, we demonstrated that we could get reasonable performance improvements with a workload that a conservative ratio of reads to writes.

\name~ will only perform well in an environment where the newest data is prioritized and compressible.
Most topologies cannot fully take advantage of \name~ unless they are built on a fog network.
Also, to truly see the best results, one must build the system on top of a dependable data-efficient backing store.
Google's API was particularly inefficient when performing reads, such that Google's Sheets API did not allow for data querying directly.
Google's API sent the entire data set over the wire if a node requested the data.

% Shortcomings:
% \begin{itemize}
% \item Context matters: performance is super workload dependent
% \item Only works in fog of a bunch of devices that support it.
% \item Results depend on efficiency of the backing system.
% Google was a particularly difficult to work with because when we read from Google we have to read out the entire table, which makes results look better.
% \end{itemize}

\section{Related Work}

Some work has been done trying to compress the data gathered from IoT nodes~\cite{datareduction,bandwidthusage,panda}.
Remote Control Caching~\cite{remotecontrolcaching} and others have implemented middlebox-based software caches for web caching.
These techniques selectively save or discard information from their sensors to reduce energy consumption while retaining the general trends in the data.

Our work draws on existing CPU cache architectures~\cite{cachecoherence,3Cs,zerocycleloads}.
Machine learning in the fog and at the edge has become an emerging area of interest in IoT~\cite{fogcomputingapps,videoanalytics}.
Some of these systems have attempted to implement distributed resource pooling, but existing implementations are controlled by a central server in the cloud.
\name~ could be deployed on edge and fog-based machine learning applications as a distributed and fault-tolerant alternative that does not require any centralized control.

\section{Conclusions}

\name~ is one possible direction in addressing the emerging problem of distributed data processing in edge and fog systems.
Furthermore, \name~ is fault tolerant---if the backing store fails, the system can recover and go on operating in steady-state as long as the uncached data is not mission-critical.
\name~ uses soft cache coherence to synchronize data among nodes in a fog.
We have demonstrated that it can reduce latency and bandwidth required to read and write recently-acquired sensor measurements by as much as 50\% compared to compared to direct writes to the backing store.
We expect performance improvements to be closely tied to the characteristics of the workload.
To understand how well distributed caches will perform in the wild, we need to evaluate \name~ with more diverse workloads on real IoT devices.
Nevertheless, \name~ is a promising technique to reduce latency and network traffic in the fog.

\bibliography{main.bib}
\bibliographystyle{IEEEtran}

\end{document}